\documentclass[twoside]{article}
\usepackage{Proc_NTSE_14}
\usepackage{bm}       % bold math 
\usepackage{amsmath}
\usepackage{graphicx,color}

\begin{document}

\newcommand {\nc} {\newcommand}
\nc {\IR} [1]{\textcolor{red}{#1}}
\nc {\IB} [1]{\textcolor{blue}{#1}}
\nc {\IM} [1]{\textcolor{magenta}{#1}}

\thispagestyle{plain}
%Please use the command \publref{myfilename} to print the reference to your proceedings contribution 
%at the bottom of the page where myfilename should be replaced by the name of your LaTeX file
%(e.g., use the command \publref{Johns} if the LaTeX file of your contribution called Johns.tex):
\publref{elster-ntse}

\begin{center}
{\Large \bf \strut
%Insert the title of your contribution here
Separable Optical Potentials for (d,p) Reactions
\strut}\\
\vspace{10mm}
{\large \bf 
%Insert the authors here. Use upper indexes a, b, c, etc., to bind authors with their addresses
% as shown below.
Ch.~Elster$^{a}$, L.~Hlophe$^{a}$, V.~Eremenko$^{a,e}$, F.M.~Nunes$^{b}$,
G.~Arbanas$^{c}$, J.E.~Escher$^d$, I.J.~Thompson{$^d$} \\
\vspace{2mm}
 (The TORUS Collaboration)}
\end{center}

\noindent{% Insert the addresses  here.
\small $^a$\it Institute of Nuclear and Particle Physics,  and
Department of Physics and Astronomy, \\  Ohio University, Athens, OH 45701, USA} \\
{\small $^b$\it National Superconducting Cyclotron Laboratory and Department of Physics
and Astronomy, Michigan State University, East Lansing, MI 48824, USA } \\
{\small $^c$\it Reactor and Nuclear Systems Division, Oak Ridge National Laboratory, \\
Oak Ridge,  TN 37831, USA} \\
 {\small $^d$\it Lawrence Livermore National Laboratory L-414, Livermore, CA 94551, USA} \\
{\small $^e$\it D.V. Skobeltsyn Institute of Nuclear Physics, M.V. Lomonosov
Moscow State University, Moscow, 119991, Russia }

%The next command defines running titles:
\markboth{
%Put here the list of authors that will be displayed in running titles:
The TORUS Collaboration}
{Separable Optical Potentials ...
} 

\begin{abstract}
An important ingredient for applications of nuclear physics to e.g. astrophysics or
nuclear energy are the cross sections for reactions of neutrons with rare isotopes. Since
direct measurements are often not possible, indirect methods like
$(d,p)$ reactions must be used instead. Those
$(d,p)$ reactions may be viewed as  effective three-body reactions and
described with Faddeev techniques.
An additional challenge posed by $(d,p)$ reactions involving
heavier nuclei is the treatment of the Coulomb force. To avoid numerical complications in
dealing with the screening of the Coulomb force, recently a new approach using the Coulomb
distorted basis in momentum space was suggested. In order to implement this suggestion, one
needs not only to derive a separable representation of neutron- and proton-nucleus optical
potentials, but also compute the Coulomb distorted form factors in this basis. 
\\[\baselineskip] 
\end{abstract}

\section{Introduction}
\label{intro}

Nuclear reactions are an important probe to learn about the structure of unstable nuclei.
Due to the short lifetimes
involved, direct measurements are usually not possible. Therefore indirect measurements using
($d,p$) reactions have been proposed (see e.g.
Refs.~\cite{RevModPhys.84.353,jolie,Kozub:2012ka}). 
Deuteron induced reactions are particularly attractive from an experimental perspective,
since deuterated targets are readily available. From a theoretical perspective they are
equally attractive because the scattering problem can be reduced to an effective three-body
problem~\cite{Nunes:2011cv}. Traditionally deuteron-induced single-neutron transfer
($d,p$) reactions have been used to study the shell structure in stable nuclei, nowadays
experimental techniques are available to apply the same approaches to exotic beams (see e.g.
\cite{Schmitt:2012bt}).
Deuteron induced $(d,p)$ or $(d,n)$ reactions in inverse kinematics are
also useful to extract neutron or proton capture rates on unstable nuclei of astrophysical
relevance. Given the many ongoing experimental programs  worldwide using these reactions, a
reliable reaction theory for $(d,p)$ reactions is critical.

One of the most challenging aspects of solving the three-body problem for nuclear reactions is
the repulsive Coulomb interaction.
While the Coulomb interaction for light nuclei is often a small correction to the problem,
this is certainly not the case for intermediate mass and heavy systems. Over the last decade,
many theoretical efforts have focused on advancing the theory
for $(d,p)$ reactions (e.g.  \cite{Mukhamedzhanov:2012qv,Deltuva:2013jna}) 
and testing existing methods (e.g.  \cite{Deltuva:2007gj,Nunes:2011cv,Upadhyay:2011ta}).  
Currently, the most complete implementation
of the theory is provided by the Lisbon group \cite{Deltuva:2009fp}, which solves the Faddeev
equations in the Alt, Grassberger and Sandhas \cite{ags} formulation. The method introduced in
\cite{Deltuva:2009fp} treats the Coulomb interaction with a screening and renormalization
procedure as detailed in  \cite{Deltuva:2005wx,Deltuva:2005cc}. While the current
implementation of the Faddeev-AGS equations with screening is computationally effective for
light systems, as the charge of the nucleus increases technical difficulties arise in the
screening procedure \cite{hites-proc}. Indeed, for most of the new exotic nuclei to be
produced at the Facility of Rare Isotope Beams, the current method is not adequate. Thus one
has to explore solutions to the nuclear reaction three-body problem where the Coulomb problem
is treated without screening.

In Ref.~\cite{Mukhamedzhanov:2012qv}, a three-body theory for $(d,p)$ reactions is derived with
explicit inclusion of target excitations, where no screening of the Coulomb force is
introduced. Therein, the Faddeev-AGS equations are cast in a Coulomb-distorted partial-wave
representation, instead of a plane-wave basis.
This approach assumes the interactions in the two-body subsystems to be separable.
While in  Ref.~\cite{Mukhamedzhanov:2012qv} the lowest angular momentum in this basis 
($l=0$) is derived for a
Yamaguchi-type nuclear interaction is derived as analytic expression,
it is desirable to implement more general form factors, which are modeled after the nuclei
under consideration.

In order to bring the three-body theory laid out in Ref.~\cite{Mukhamedzhanov:2012qv} to
fruition, well defined  preparatory work needs to be successfully carried out. 
Any momentum space Faddeev-AGS type calculation needs as input transition matrix elements in
the different two-body subsystems. In the case of ($d,p$) reactions with nuclei these are
the $t$-matrix elements obtained from the neutron-proton, the neutron-nucleus and
proton-nucleus interactions. Since the formulation in Ref.~\cite{Mukhamedzhanov:2012qv}
is designed for separable interactions, those need to be developed not only in the
traditionally employed plane wave basis, but also the basis of Coulomb scattering states.

This contribution summarizes the three major developments required to provide
reliable input to the three-body formulation for $(d, p)$ reactions 
without screening the Coulomb force, namely
\begin{itemize}
\item the derivation of momentum-space separable representations of 
  neutron-nucleus optical potentials~\cite{Hlophe:2013xca},
\item  the derivation of momentum-space separable representations of
  proton-nucleus optical potentials in the Coulomb basis~\cite{Hlophe:2014soa},
\item the calculation of neutron-nucleus form-factors in
  the basis of momentum-space Coulomb scattering states~\cite{upadhyay:2014}.
\end{itemize}
Sections~\ref{nplusA}, \ref{pplusA}, and~\ref{cnplusA} summarize
the necessary steps to achieve reliable calculations of those input quantities 
needed for  three-body calculations that treat the Coulomb force without screening.
Finally, we summarize in Section~\ref{summary}.

%%%%%%%%%%%%%%%%%%%%%%%%%%%%%%%%%%%%%%%%%%%%%%%%%%%%%%%%%%%%%%%%%%%%%%%%%%%%%%%%%%

\section{Separable Representation of Nucleon-Nucleus \\ Optical Potentials}
\label{nplusA}

Separable representations of the forces between constituents forming the subsystems in a
Faddeev approach have a long tradition in few-body physics. There is a large body of work on
separable representations of
nucleon-nucleon (NN) interactions (see e.g.
Refs.~\cite{Haidenbauer:1982if,Haidenbauer:1986zza,Berthold:1990zz,Schnizer:1990gf,Entem:2001it})
or  meson-nucleon interactions~\cite{Ueda:1994ur,Gal:2011yp}.
In the context of describing light nuclei like $^6$He~\cite{Ghovanlou:1974zza}
and $^6$Li~\cite{Eskandarian:1992zz} in a three-body approach, separable interactions have
been successfully used.
A separable nucleon-$^{12}$C optical potential was proposed in
Ref.~\cite{MiyagawaK}, consisting of a rank-1 Yamaguchi-type form factor fitted to
the positive energies and a similar term describing the bound states in the
nucleon-$^{12}$C configuration.  However, 
systematic work along this line for heavy nuclei,
for which  excellent phenomenological descriptions exist in terms of
Woods-Saxon functions~\cite{Varner:1991zz,Weppner:2009qy,Koning:2003zz,Becchetti:1969zz} has
not been carried out until recently~\cite{Hlophe:2013xca}.

The separable representation of two-body interactions suggested by
Ernst-Shakin-Thaler~\cite{Ernst:1973zzb} (EST) is well suited for achieving
this goal. We note that this EST approach
has been successfully employed to represent NN
potentials~\cite{Haidenbauer:1982if,Haidenbauer:1986zza}.
However, the EST scheme derived in Ref.~\cite{Ernst:1973zzb},
 though allowing energy dependence of the potentials~\cite{Ernst:1974zzb,Pearce:1987zz},
assumes that they are Hermitian.
Therefore, we generalized the EST approach in Ref.~\cite{Hlophe:2013xca} 
in order to be applicable for optical potentials which are complex. 
For the ease of the reader, we briefly summarize the main points of that work. 

For applications to the theory of nuclear reactions  all
potential operators $U$ need to satisfy
\begin{equation}
{\cal K} U {\cal K}^{-1} = U^\dagger,
\label{eq:2.1}
\end{equation}
where $\cal{K}$ is the time reversal operator appropriate to the system. This condition
guarantees that the $S$-matrix corresponding to $U$ is symmetric  and that reaction
amplitudes constructed from these potentials satisfy reciprocity relations.
When  $U$ is a central potential in the space of a spinless particle,  $\cal{K}$ can be
chosen
to be the anti-linear complex conjugation operator ${\cal K}_0$, which in the
coordinate space basis $|{\bf r}\rangle$ is
 defined by
\begin{equation}
{\cal K}_0\; \alpha \,|{\bf r}\rangle = \alpha^* ( {\cal K}_0 |{\bf r}\rangle) =
\alpha^*|{\bf r}\rangle,
\label{eq:2.2}
\end{equation}
and from which we deduce ${\cal K}_0 |{\bf p}\rangle = |-{\bf p}\rangle.$
 Note that for this particular  ${\cal K}$ we have $({\cal K}_0)^{-1}={\cal K}_0$.

Considering first a rank-1 separable potential, the EST scheme presented in 
Ref.~\cite{Ernst:1973zzb} requires that a separable potential ${\bf U}$ leads to the same
scattering wave functions at a specific energy $E_{k_E}$  (support point) as the
potential $u$ it is supposed to represent.  For $u$ being a non-Hermitian potential, 
we define
\begin{equation}
{\bf U}(E_{k_E}) \equiv \frac{ u | f_{l,k_E}\rangle \langle  f^*_{l,k_E}|u}
           {\langle f^*_{l,k_E} |u| f_{l,k_E}\rangle} \equiv u |
f_{l,k_E}\rangle {\hat \lambda} \langle f^*_{l,k_E}|u \; ,
\label{eq:2.4}
\end{equation}
where the strength parameter is defined by $({\hat \lambda})^{-1} = \langle
f^*_{l,k_E} |u| f_{l,k_E}\rangle $.
Here $f_{l,k_E}(r)$ is the unique regular radial wave function corresponding to $u$ and
$f^*_{l,k_E}(r)$ is the unique regular radial wavefunction corresponding to $u^*$. By a
suitable choice of arbitrary normalization constants we can arrange that
$f^*_{l,k_E}(r)$ is simply the complex conjugate of $f_{l,k_E}$ and hence ${\cal
K}_0|f_{l,k_E}\rangle=|f^*_{l,k_E}\rangle $.

If $u$ satisfies ${\cal K}_0 u {\cal K}_0=u^{\dagger}$ the definition of
Eq.~(\ref{eq:2.4})
gives  a symmetric complex potential matrix that satisfies
\begin{equation}
{\cal K}_0{\bf U}(E_{k_E}){\cal K}_0 =
  \left[ \vphantom{f^*_{l,k_E}} {\cal K}_0 u | f_{l,k_E}\rangle \right]
  ({\hat \lambda})^*
  \left[ \langle f^*_{l,k_E}|u {\cal K}_0 \right]
= u^{\dagger} | f^*_{l,k_E}\rangle
  ({\hat \lambda})^*
  \langle f_{l,k_E}|u^{\dagger}
= U^{\dagger},
\label{eq:2.5}
\end{equation}
where the square brackets mean that ${\cal K}_0$ here acts only on the quantities within
the brackets.

In analogy to the procedure followed in Ref.~\cite{Ernst:1973zzb} we
define a complex separable potential of arbitrary rank in a given partial wave
as
\begin{equation}
{\bf U}  = \sum_{i,j} u| f_{l,k_{E_i}} \rangle \langle f_{l,k_{E_i}} |M |
f^*{_l,k_{E_j}}\rangle
\langle f^*_{l,k_{E_j}}|u .
\label{eq:2.6}
\end{equation}
Here $f_{l,k_{E_i}}$ and $f^*_{l,k_{E_i}}$  are the same unique regular radial wave function 
as used in Eq.~(\ref{eq:2.4}). Note that $u$ may also be energy dependent.

\begin{figure}[h]
\centerline{\includegraphics[width=10.0cm]{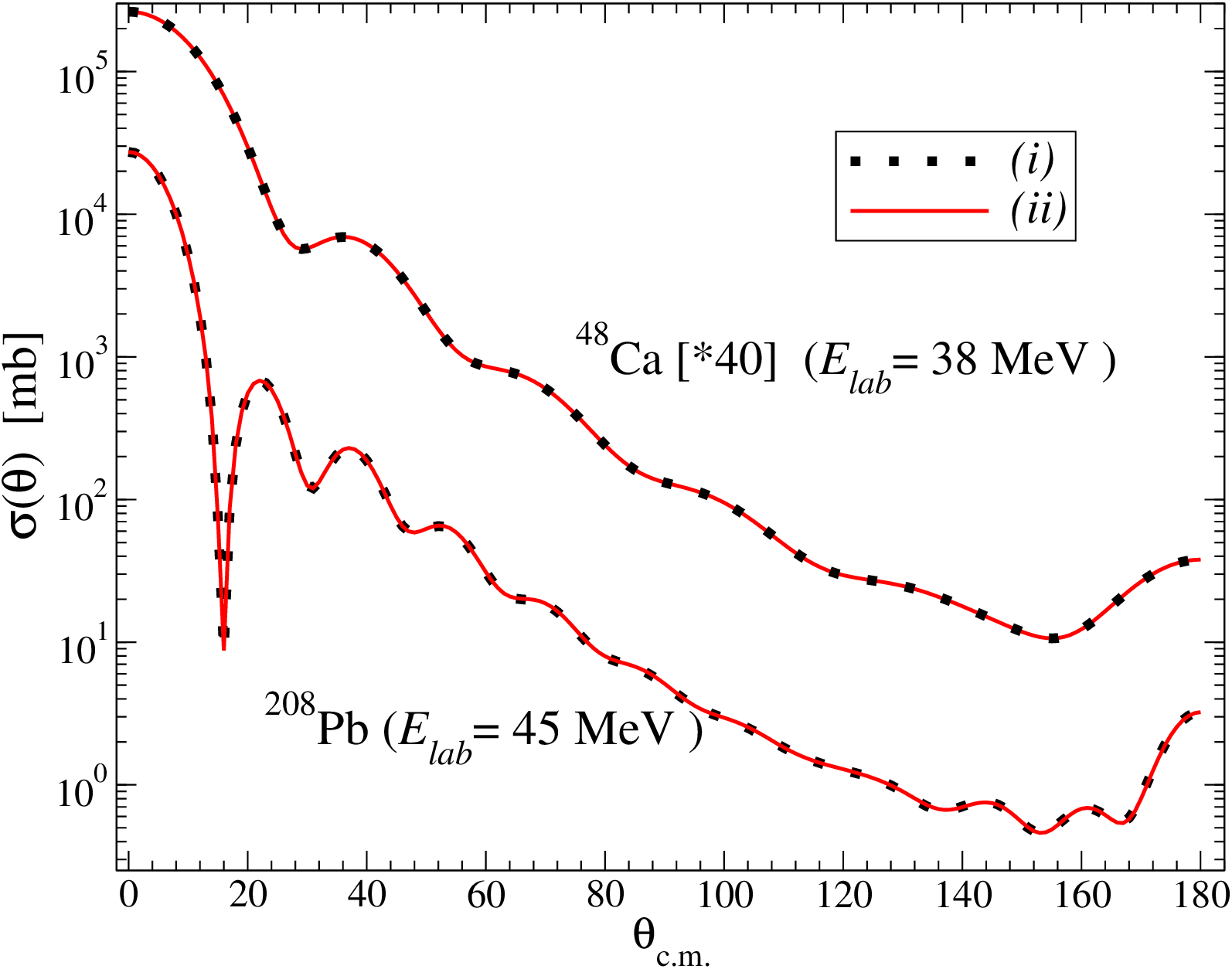}}
\caption{The unpolarized differential cross section for elastic scattering of 
neutrons from $^{48}$Ca (upper) and $^{208}$Pb (lower) as function of the 
c.m. angle. For $^{48}$Ca the cross section is calculated at a
laboratory kinetic energy of 38~MeV and is scaled by a factor 40. The calculation for
$^{208}$Pb is carried out at $E_{lab}$~=~45~MeV. The solid lines ($ii$) depict the cross
section calculated in momentum space based on the rank-5 separable representation of the
CH89~\cite{Varner:1991zz} phenomenological optical potential, while the dotted lines
($i$) represent the corresponding coordinate space calculations.
}
\label{fig1}
\end{figure}

\noindent
The  matrix $M$ is  defined and constrained by
\begin{eqnarray}
\delta_{ik}&=&\sum_j \langle f_{l,k_{E_i}}|M| f^*_{l,k_{E_j}}\rangle \langle
f^*_{l,k_{E_j}} | u| f_{l,k_{E_k}}\rangle  \cr
&=& \sum_j \langle f^*_{l,k_{E_i}} | u | f_{l,k_{E_j}} \rangle \langle f_{l,k_{E_j}}
|M|f^*_{l,k_{E_k}}\rangle.
\label{eq:2.7}
\end{eqnarray}
The corresponding separable partial wave $t$-matrix must be of the form
\begin{equation}
t(E) =  \sum_{i,j} u| f_{l,k_{E_i}}\rangle \tau_{ij}(E) \langle f^*_{l,k_{E_j}}|u \; ,
\label{eq:2.8}
\end{equation}
with the following  restrictions
\begin{align}
\delta_{nj} &=
  \sum_i \langle f^*_{l,k_{E_n}}|u - u g_0(E) u| f_{l,k_{E_i}}\rangle
  \; \tau_{ij}(E),\label{eq:2.9}\\
\delta_{ik} &=
  \sum_j \tau_{ij}(E) \; \langle f^*_{l,k_{E_j}}|u -u g_0(E) u
  | f_{l,k_{E_k}}\rangle.\label{eq:2.10}
\end{align}
For the explicit calculation of the matrix   $\tau_{ij}(E)$, we define a matrix
\begin{equation}
R_{ij}(E) \equiv \langle f^*{_l,k_{E_i}}|u -u g_0(E) u| f_{l,k_{E_j}}\rangle ,
\label{eq:2.11}
\end{equation}
so that the  condition of Eq.~(\ref{eq:2.10}) reads
\begin{equation}
\sum_j \tau_{ij}(E) R_{jk}(E) = \delta_{ik},
\label{eq:2.12}
\end{equation}
from which follows
\begin{equation}
\tau_{ij} (E) = \left( R(E) \right)^{-1}_{ij}.
\label{eq:2.13}
\end{equation}
Using  that
$ t (p',k_{E_i}, E_{i})=\langle f^*_{l,k_{E_i}}|u| p' \rangle$, and
$ t (p,k_{E_i}, E_{i})=\langle p |u| f_{l,k_{E_i}} \rangle$,
the matrix elements $R_{ij}$ are calculated  in momentum space as given explicitly in
Ref.~\cite{Hlophe:2013xca}.

In order to demonstrate the construction of a separable representation of a complex
potential we apply the generalized EST scheme to neutron scattering from $^{48}$Ca and
$^{208}$Pb and use as starting point the Chapel Hill phenomenological global optical
potential CH89~\cite{Varner:1991zz}, which has been widely used in the literature over
the last decades. Like most phenomenological global optical potentials,
CH89 is based on
Woods-Saxon functions, which are more naturally given in coordinate space, and have an
explicit energy dependence in the strength functions. In order to derive a separable
momentum-space representation of CH89, we first must construct a momentum-space
representation of the potential itself. The Fourier transform of Woods-Saxon functions
leads to a series expansion in momentum space, of which only the first two terms
are necessary to obtain a converged result~\cite{Hlophe:2013xca}. The momentum-space
potential then enters a Lippmann-Schwinger (LS) integral equation to obtain the
half-shell t-matrices at fixed energies (support points) $E_i$, from which the separable
representation given in  Eq.~(\ref{eq:2.8}) is then constructed after having obtained
the coupling matrix $\tau_{ij}(E)$ from the solution of Eq.~(\ref{eq:2.13}).  

\begin{table}
\caption{The EST support points at c.m. energies $E_{k_i}$
used  for constructing the separable
representation of the partial wave s-matrix of the n$+^{48}$Ca
 and n$+^{208}$Pb systems.
The support points in the last row for the n$+^{208}$Pb system
given in bold face indicate the universal set of support points, which can be used to
construct a representation for all nuclei given by the
CH89~\cite{Varner:1991zz} phenomenological optical potential.
}
\label{table1}
\begin{center}
\begin{tabular}{|c|c|c|c| }
\hline\hline
   system    & partial wave(s)  & rank & EST support point(s) [MeV]\\
\hline
             &  $l\ge 10$       &  1   & 40         \\
n$+^{48}$Ca  &  $l\ge  8$       &  2   & 29, 47      \\
             &  $l\ge  6$       &  3   & 16, 36, 47      \\
             &  $l\ge  0$       &  4   & 6,  15, 36, 47   \\[1ex]
\hline
%             &                   &      &                \\
             &  $l\ge  16$       &  1   & 40 \\
n$+^{208}$Pb &  $l\ge  13$       &  2   & 35, 48   \\
            &  $l\ge  11$       &  3   & 24, 39, 48       \\
            &  $l\ge   6$       &  4   & 11, 21, 36, 45          \\
          &  $\mathbf {l\ge   0}$       &  $\mathbf{5}$   & $\mathbf{ 5, 11, 21, 36, 47}$
\\[1ex]
\hline
\end{tabular}
\end{center}
\end{table}

A major finding of Ref.~\cite{Hlophe:2013xca} is a systematic classification of support
points for partial wave groups, so that the partial wave S-matrix elements are reproduced
to at least four significant figures compared to the original momentum-space solution of
the LS equation. It turns out that the low partial waves of the $n+^{208}$Pb system
require a rank-5 separable potential to be well represented in the energy regime between
0 and 50~MeV center-of-mass energy. The rank required for achieving a good
representation decreases with increasing angular momentum of the partial wave
considered. The recommendation of Ref.~\cite{Hlophe:2013xca} for both the rank and the
locations of the support points to be used when describing medium-mass and heavy systems
generated from the CH89 potential are repeated in Table~\ref{table1} for the convenience
of the reader. 

In order to demonstrate the quality of the separable representations obtained with 
the generalized EST scheme, Fig.~\ref{fig1}
depicts the unpolarized differential cross section for elastic scattering of neutrons
from $^{48}$Ca at 38~MeV laboratory kinetic energy and from $^{206}$Pb at 45~MeV as
function of the center-of-mass (c.m.) angle $\theta_{c.m.}$. The
solid lines ($i$) represent the calculations with the separable representations, while 
the dotted lines ($ii$) stand for the corresponding coordinate space calculations. The
agreement is excellent over the entire angular range, indicating that all partial wave
S-matrix elements that enter the cross section are well described by the separable
representation. 

%%%%%%%%%%%%%%%%%%%%%%%%%%%%%%%%%%%%%%%%%%%%%%%%%%%%%%%%%%%%%%%%%%%%%%%%%%%%%%%%%%%%

\section{Separable Representation of Proton-Nucleus \\ Optical Potentials
in the Coulomb Basis}
\label{pplusA}

In order to implement the formulation of the Faddeev-AGS equations proposed in
Ref.~\cite{Mukhamedzhanov:2012qv} we need the proton-nucleus form factors in the
Coulomb distorted basis,
and thus need to have a separable representation of proton-nucleus optical potentials.
In Refs.~\cite{Cattapan:1975tf,Cattapan:1975np}
rank-1 separable interactions of Yamaguchi form were introduced to represent the
nuclear force up to a few MeV, and the Coulomb distorted basis was
introduced to compute proton elastic scattering from light nuclei. This is
not sufficient for considering the proton-nucleus interaction in a
separable representation for scattering of heavy nuclei up to tens of MeV.
Thus we need to extend the generalization of the EST scheme presented in the previous
section such that it can be applied in the Coulomb distorted basis~\cite{Hlophe:2014soa}. 

In general the scattering between a proton and a nucleus is governed by a potential
\begin{equation}
w = v^c + u^s ,
\label{eq:0a}
\end{equation}
where $v^c$ is the repulsive Coulomb potential and $u^s$ an arbitrary short range
potential.
In general $u^s$ consists of an optical potential, which describes the nuclear
interactions and a short-ranged Coulomb potential traditionally parameterized 
as the potential of a charged sphere
with radius $R_0$ from which the point Coulomb force is
subtracted~\cite{Varner:1991zz}. In practice, 
\begin{equation}
u^s = u^N +(v^{cd} - v^c),
\label{eq:3.1}
\end{equation}
where $u^N$ represents the nuclear (optical) potential,  $v^{cd}$ is the Coulomb potential 
inside the nucleus, and is usually taken as the
Coulomb potential for a uniformly charged sphere of radius $R_0$, from which the point
Coulomb potential is subtracted. The expressions for the short-ranged charge
distribution is given in Ref.~\cite{Varner:1991zz} as
\begin{equation}
(v^{cd} - v^c)(r) = \alpha Z_1 Z_2 \left[ \frac{1}{2R_0} \left( 3 -
\frac{r^2}{R_0^2} \right) -
\frac{1}{r} \right],
\label{eq:3.2}
\end{equation}
with  $Z_1$ and $Z_2$
being the atomic numbers of the particles, and $\alpha$ the Coulomb coupling constant.
Since the scattering problem governed by the point Coulomb force has
an analytic solution,
the scattering amplitude for elastic scattering between a
proton and a spin-zero nucleus is obtained as the sum of the Rutherford
amplitude $f^C(E_{p_0},\theta)$  and the Coulomb distorted nuclear amplitude given by
\begin{equation}
M^{CN} (E_{p_0},\theta) = f^{CN}(E_{p_0},\theta)+{\hat\sigma}\cdot
\mathbf{\hat{n}}\;  g^{CN}(E_{p_0},\theta), 
\end{equation}
with
\begin{eqnarray}
\lefteqn{f^{CN}(E_{p_0},\theta)=} & &  \\
&& -\pi\mu \sum_{l=0}^\infty e^{2i\sigma_l(E_{p_0})} P_l(\cos \theta) \times 
 \Big[ (l+1) \langle p_0  | \tau^{CN}_{l+}(E_{p_0})| p_0 \rangle + l
\langle p_0|\tau^{CN}_{l-}(E_{p_0})| p_0 \rangle\Big] , \nonumber 
\end{eqnarray}
and
\begin{eqnarray}
\lefteqn{ g^{CN}(E_{p_0},\theta) =}&  &  \\
&& -\pi\mu \sum_{l=0}^\infty e^{2i\sigma_l(E_{p_0})} P^1_l(\cos \theta) \times 
 \Big[\langle p_0| \tau^{CN}_{l+}(E_{p_0})| p_0 \rangle - \langle
p_0|\tau^{CN}_{l-}(E_{p_0})| p_0 \rangle\Big] . \nonumber
\label{eq:3.3}
\end{eqnarray}
Here $E_{p_0}=p_0^2/2\mu$ is the center-of-mass (c.m.) scattering energy which defines
the on-shell momentum $p_0$, and $\sigma_l=\;\arg\Gamma(1+l+i\eta)$ is the Coulomb phase
shift.  The Sommerfeld parameter is given by $\eta=\alpha Z_1Z_2\mu/p_0$.
The unit vector $\mathbf{\hat{n}}$ is normal to the scattering plane, and
${\hat\sigma}/2$
is the spin operator. The subscripts $'+'$ and $'-'$ correspond to a total angular
momentum $j=l+1/2$ and $j=l-1/2$.  

Suppressing the total angular momentum indices for simplicity,
the Coulomb distorted nuclear $t$-matrix element is given by
$\langle p_0|\tau^{CN}_l(E_{p_0})| p_0 \rangle$, which is the solution of a LS type
equation,
\begin{multline}
\langle p | \tau^{CN}_l(E_{p_0})| p_0 \rangle  = \langle p | u^s_l |
  p_0 \rangle \\
+ \int p'^2 dp'  \langle p | u^s_l | p'\rangle \langle p'| g_c
(E_{p_0} +i\varepsilon)|p' \rangle \langle p' |\tau^{CN}_l(E_{p_0})| p_0
\rangle.
\label{eq:3.4}
\end{multline}
Here
\begin{equation}
g_c^{-1}(E_{p_0} +i\varepsilon)  = E_{p_0} + i\varepsilon  -H_0 - v^c
\label{eq:3.5}
\end{equation}
 is the Coulomb
Green's function and  $H_0$  the free Hamiltonian.
The Coulomb distorted nuclear $t$-matrix element $\langle p | \tau^{CN}_l(E_{p_0})| p_0
\rangle$
is related to the proton-nucleus $t$-matrix $\langle p|t_l(E_{p_0})|p_0\rangle$ by the
familiar two-potential formula
\begin{eqnarray}
 \langle p|t_l(E_{p_0})|p_0\rangle =
\langle p|t^C_l(E_{p_0})|p_0\rangle+
 e^{2i\sigma_l(E_{p_0})}\langle p | \tau^{CN}_l(E_{p_0})| p_0 \rangle,
 \label{eq:3.6}
\end{eqnarray}
where $\langle p|t^C_l(E_{p_0})|p_0\rangle$ is the point Coulomb $t$-matrix.
When the integral equation, Eq.~(\ref{eq:3.4}), is solved in the basis of
Coulomb eigenfunctions,
$g_c$ acquires the form of a free Green's function and the difficulty of
solving it is shifted to evaluating the potential matrix
elements in this basis.
For deriving a separable representation of the Coulomb distorted proton-nucleus
$t$-matrix element,
we generalize the approach suggested by Ernst, Shakin, and Thaler
(EST)~\cite{Ernst:1973zzb},
 to the charged particle case. The basic idea behind the EST
construction of a separable representation of a given potential is that
the wave functions
calculated with this potential and the corresponding separable potential agree at given
fixed scattering energies $E_i$, the EST support points. The formal derivations
of~\cite{Ernst:1973zzb} use the plane wave basis, which is standard for scattering
involving short-range potentials. However, the EST scheme does not depend on the basis and can
equally well be carried out in the basis of Coulomb scattering wave functions.

In order to generalize the EST approach to charged-particle scattering,
one needs to be able to obtain the scattering wave functions or half-shell
t-matrices from a given potential in the Coulomb basis, and then construct the
corresponding separable representation thereof.

In order to calculate the  half-shell $t$-matrix of Eq.~(\ref{eq:3.3}),
we evaluate the integral
equation in the Coulomb basis as suggested in~\cite{Elster:1993dv} and successfully
applied in~\cite{Chinn:1991jb}, and note that
in this case the Coulomb Green's function behaves like a free Green's function.
Taking $|\Phi_{l,p}^{c}\rangle$ to represent the partial wave Coulomb eigenstate,
the LS equation becomes
\begin{eqnarray}
\langle \Phi_{l,p}^{c} |\tau^{CN}_l(E_{p_0})|\Phi_{l,p_0}^{c}\rangle &=&
\langle \Phi_{l,p}^{c} | u^s | \Phi_{l,p_0}^{c}\rangle  +  \cr
& \int\limits_0^\infty & \langle \Phi_{l,p}^{c} |u^s| \Phi_{l,p'}^{c} \rangle
\;\frac { p'^2 dp'}{E_{p_0} - E_{p'} +i\varepsilon} \langle
\Phi_{l,p'}^{c}|\tau^{CN}_l(E_{p_0})|\Phi_{l,p_0}^{c}\rangle  \cr
& \equiv & \langle p | \tau^{CN}_l(E_{p_0})| p_0 \rangle , 
\label{eq:3.7}
\end{eqnarray}
which defines the Coulomb distorted nuclear $t$-matrix of Eq.~(\ref{eq:3.4}).

\begin{figure}
\centerline{\includegraphics[width=9.1cm]{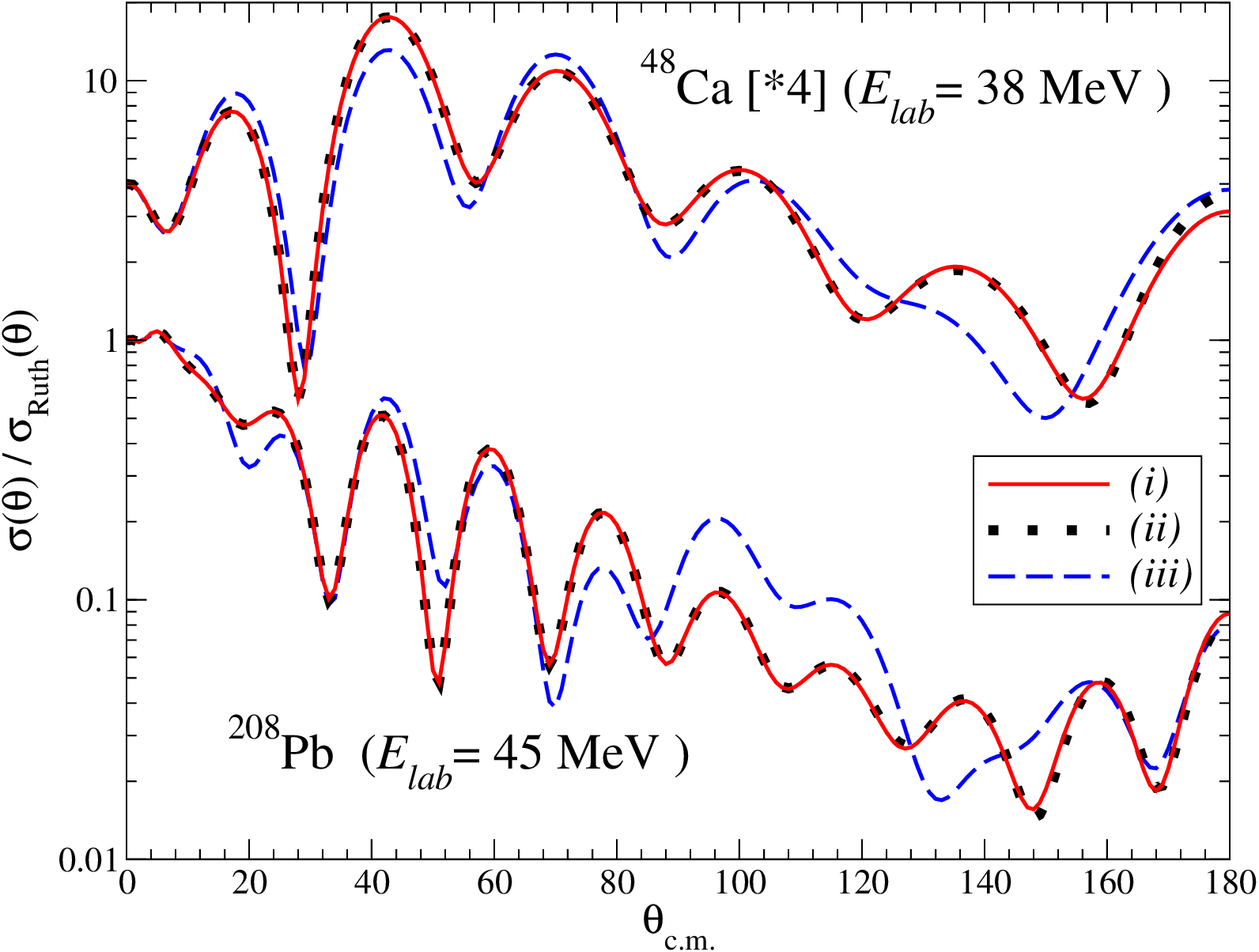}}
\caption{The unpolarized differential cross section for elastic scattering of 
protons from $^{48}$Ca (upper) and $^{208}$Pb (lower) divided by the Rutherford cross
section as function of the 
c.m. angle $\theta_{c.m.}$. For $^{48}$Ca the cross section is calculated at a
laboratory kinetic energy of 38~MeV and is scaled by a factor 4. The calculation for
$^{208}$Pb is carried out at $E_{lab}$~=~45~MeV. The solid lines ($i$) depict the cross
section calculated in momentum space based on the rank-5 separable representation of the
CH89~\cite{Varner:1991zz} phenomenological optical potential, while the dotted lines
($ii$) represent the corresponding coordinate space calculations. The dash-dotted lines
($iii$) show calculations in which the short-ranged Coulomb potential is omitted.
}
\label{fig2}
\end{figure}

To determine the short-range potential matrix element, we follow
Ref.~\cite{Elster:1993dv} and insert a complete set of position space
eigenfunctions \newpage
\begin{eqnarray}
  \langle  \Phi_{l,p'}^{c} |u^s_{l}| \Phi_{l,p}^{c}\rangle 
&=& \frac{2}{\pi}\int\limits_0^\infty \langle\Phi_{l,p'}^{c}|r' \rangle
  \;r'^2dr' \;\langle r'|u^s_{l}|r\rangle  \;r^2 dr\;\langle r| \Phi_{l,p}^{c}\rangle \cr
  & =& \frac{2}{\pi p' p} \int\limits_0^\infty r r' dr dr'  \;F_l(\eta',p'r')\;
  \langle r'| u^s_l |r \rangle  \;F_l(\eta,pr).
\label{eq:3.8}
\end{eqnarray}
The partial wave Coulomb functions are given in coordinate space as
\begin{equation}
  \langle r| \Phi_{l,p}^{c}\rangle \equiv  \frac{ F_l(\eta,pr)}{pr},
  \label{eq:3.9}
\end{equation}
where $F_l(\eta,pr)$ are the standard Coulomb functions~\cite{AbramovitzStegun},
and $\eta$($\eta'$) is the Sommerfeld parameter determined with momentum $p$($p'$).

\begin{figure}
\centerline{\includegraphics[width=10.0cm]{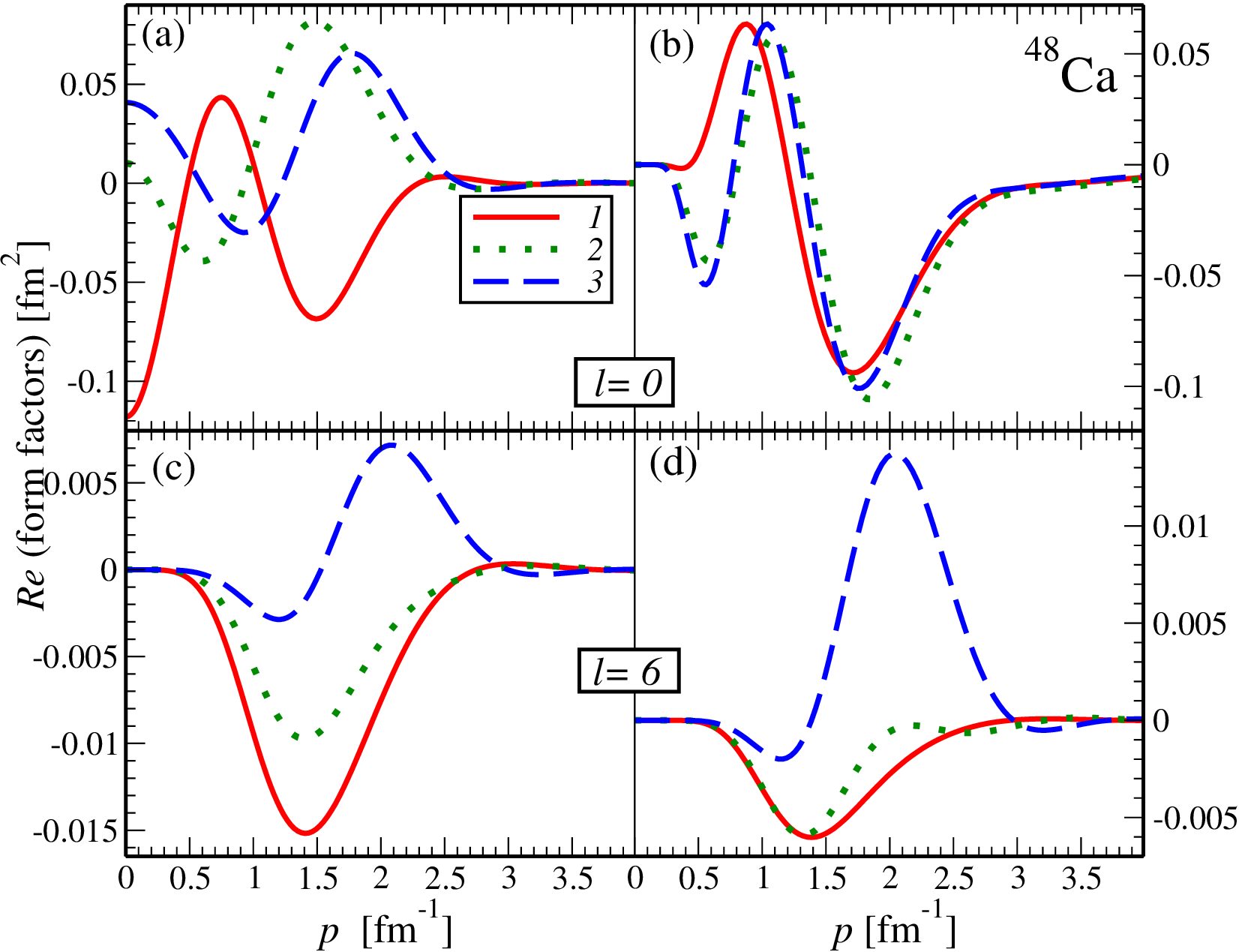}}
\caption{The real parts of the partial wave neutron form factors for $^{48}$Ca as function of
the momentum $p$ for $l=0$ (a) and $l=6$ (c). The form factors are calculated at the energies
indicated in Table~\ref{table1} for the given angular momentum, $1 \equiv$~6~MeV, $2\equiv$~15~MeV, and
$3\equiv$~36~MeV. The real parts of the proton
form factors for $^{48}$Ca as function of the momentum $p$ are given for $l=0$ in (b) and $l=6$
in (d) for the energies indicated in Table~\ref{table1}.
}
\label{fig3}
\end{figure}

\begin{figure}
\centerline{\includegraphics[width=10.0cm]{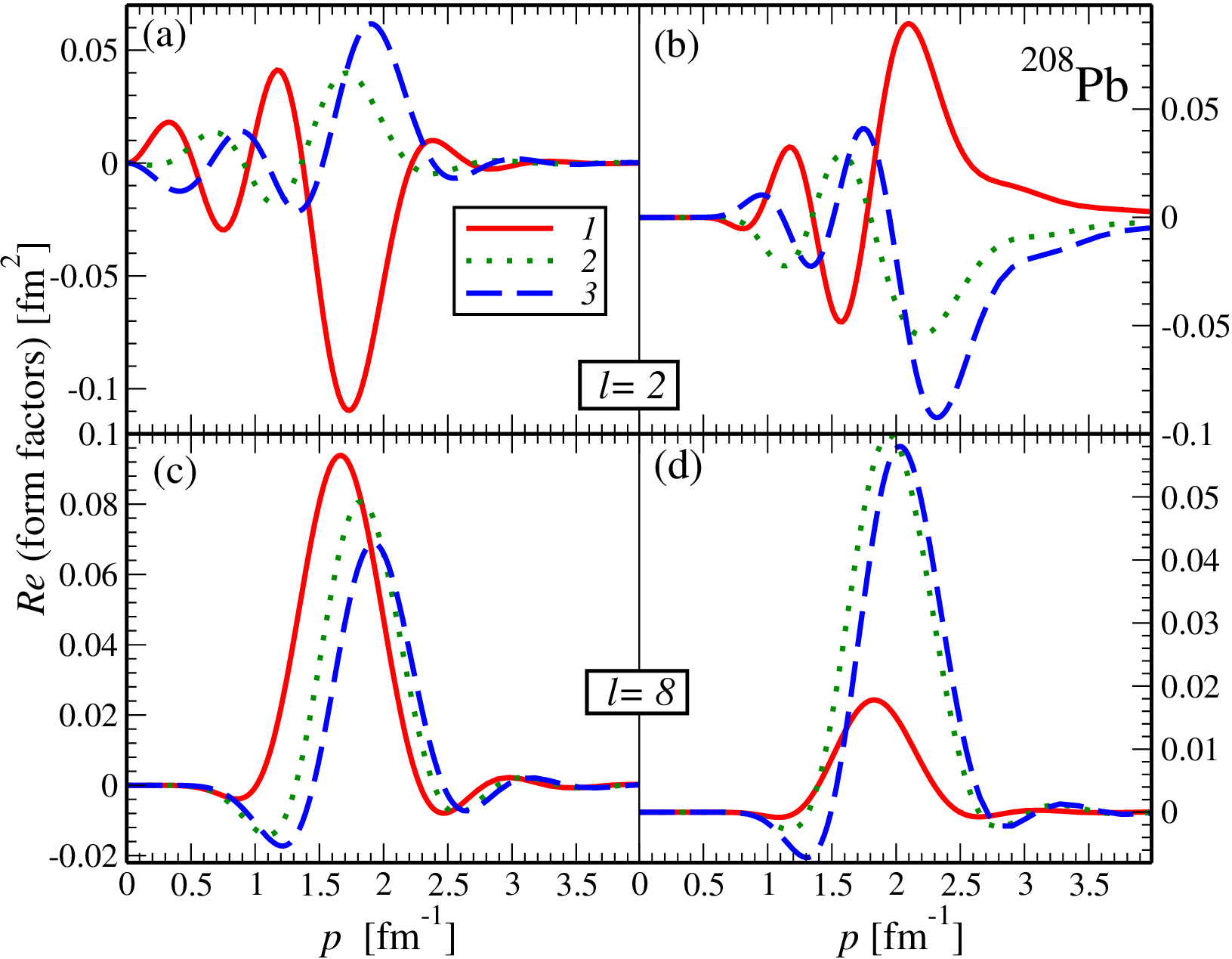}}
\caption{ The real parts of the partial wave neutron form factors for $^{208}$Pb as function of
the momentum $p$ for $l=0$ (a) and $l=8$ (c). The form factors are calculated at the first
three energies
indicated in Table~\ref{table1} for the given angular momentum, $1\equiv$~21~MeV, 
$2\equiv$~36~MeV, and $3 \equiv$~47~MeV ($l=2$) and 45~MeV ($l=8$).
The real parts of the proton
form factors for $^{208}$Pb as function of the momentum $p$ are given for $l=0$ in (b) and $l=8$
in (d) for the same energies.
}
\label{fig4}
\end{figure}

For our application we consider phenomenological optical potentials of Woods-Saxon
 form which are local in coordinate space. Thus the momentum space potential matrix
elements simplify to
\begin{eqnarray}
\langle \Phi_{l,p'}^{c} |u^s_l| \Phi_{l,p}^{c} \rangle = 
\frac{2}{\pi p'p} \int\limits_0^\infty dr\;  F_l(\eta',p' r) u^s_l(r) F_l(\eta,pr).
\label{eq:3.10}
\end{eqnarray}
We compute these matrix elements for the short-range piece of the CH89 phenomenological
global optical potential~\cite{Varner:1991zz},
which consists of the nuclear part parameterized in terms of Woods-Saxon functions and
the short-range Coulomb force of Eq.~(\ref{eq:3.2}).  The integral of Eq.~(\ref{eq:3.10}) 
can be carried out with standard methods,
since $u^s(r)$ is short ranged and the coordinate space Coulomb wavefunctions are
well defined. The accuracy of this integral can be tested by replacing the Coulomb
functions
with spherical Bessel functions and comparing the resulting matrix elements to the
partial-wave decomposition
of the semi-analytic Fourier transform used for the calculations in the previous
Section. For the cases we studied  a maximum radius of 14~fm, 300 grid points
are sufficient to obtain matrix elements with a  precision of six significant digits.

Extending the EST separable representation to the Coulomb basis involves replacing
 the neutron-nucleus half-shell $t$-matrix in Eqs.~(\ref{eq:2.7})-(\ref{eq:2.9})
 by the Coulomb distorted nuclear half-shell $t$-matrix. This leads to the
 separable Coulomb distorted nuclear $t$-matrix
\begin{equation}
 \tau^{CN}_l(E_{p_0})=\sum_{i,j} u^s|f^c_{l,k_{E_i}}\rangle \;\tau^{c}_{ij}(E_{p_0})\;
 \langle {f^c}^*_{l,k_{E_j}}|u^s ,
 \label{eq:3.11}
\end{equation}
with $\tau^c_{ij}(E_{p_0})$  being constrained by
\begin{eqnarray}
&& \sum_i \langle {f^c}^*_{l,k_{E_n}}|u^s - u^s g_c(E_{p_0}) u^s|
f^c_{l,k_{E_i}}\rangle \tau^{c}_{ij}(E) = \delta_{nj}  \\
&& \sum_j \tau^{CN}_{ij}(E_{p_0}) \; \langle {f^c}^*_{l,k_{E_j}}|u^s -u^s g_c
(E_{p_0}) u^s|
f^{c}_{l,k_{E_k}}\rangle = \delta_{ik} \;. \nonumber
\label{eq:3.12}
\end{eqnarray}
Here $| f^c_{l,k_{E_i}}\rangle$  and $| {f^c}^*_{l,k_{E_i}}\rangle$ are the regular
radial scattering wave functions corresponding to the short range potentials $u^s$
and $(u^s)^*$ at energy $E_{i}$. The separable Coulomb distorted nuclear $t$-matrix
elements are given by
 \begin{eqnarray}
 \langle p'|\tau^{CN}_l(E_{p_0})|p\rangle &\equiv& \sum_{i,j}  h_{l,i}^c(p')
\tau^c_{ij}(E_{p_0}) h_{l,j}^c(p)  \cr
& =& \sum_{i,j}\langle \Phi_{l,p'}^{c}| u^s| f^c_{l,k_{E_i}}\rangle
\tau^c_{ij}(E_{p_0}) 
\langle {f^c}^*_{l,k_{E_j}} | u^s  | \Phi_{l,p}^{c}\rangle ,
\label{eq:3.13}
\end{eqnarray}
where the form factor 
\begin{eqnarray}
h_{l,i}^c(p) &\equiv & \langle \Phi_{l,p}^{c}| u^s| f^c_{l,k_{E_i}}\rangle \\
&=& \langle {f^c}^*_{l,k_{E_i}}|u^s| \Phi_{l,p}^{c}\rangle= 
\langle p|\tau^{CN}_{l}(E_i)| k_{E_i} \rangle \nonumber
\end{eqnarray}
is the Coulomb distorted short-range
half-shell $t$-matrix satisfying Eq.~(\ref{eq:3.7}).
We want to point out that the generalization of the EST scheme to complex potentials is
not affected by changing the basis from plane waves to Coulomb scattering states.

For studying the quality of the representation of proton-nucleus optical potentials
we consider p+$^{48}$Ca and p+$^{208}$Pb elastic scattering and show the unpolarized
differential cross sections divided by the Rutherford cross section
 as function of the c.m. angle $\theta_{c.m.}$ in
Fig.~\ref{fig2}. First, we observe  very good agreement in both cases of the momentum space
calculations using the separable representation  with the corresponding coordinate 
space calculations. Second, we want to point out that we used for the separable 
representation of
the proton-nucleus partial-wave $t$-matrices the same support points (Table~\ref{table1}) 
as in the neutron-nucleus case. This makes the determination of suitable support points
$E_i$ for a given optical potential and nucleus quite efficient.  
In Fig.~\ref{fig2} we also show a calculation in which the short-range Coulomb
potential of Eq.~(\ref{eq:3.3}) is omitted. The differences in the cross sections
clearly demonstrate the importance of including this term. A detailed comparison 
of the partial-wave S-matrix elements as function of the angular momentum is given in
Ref.~\cite{Hlophe:2014soa}.

In order to illustrate some details of the separable representation of the $t$-matrix of
Eq.~(\ref{eq:2.8}) that leads to the cross section given in Fig.~\ref{fig1}, 
we display in the left panels of Fig.~\ref{fig3}  the real parts of the form factors of the 
$n+^{48}$Ca $t$-matrix for $l=0$ (a) and $l=6$ (c) at  support points given in 
Table~\ref{table1} for the respective angular momentum. 
Only for $l=0$ the form factors have a finite value at $p=0$,
while for the higher angular momentum all form factors go to zero for $p \rightarrow 0$ due to
the angular momentum barrier. For comparison, the right panels in Fig.~\ref{fig3} display
the form factors of the Coulomb distorted nuclear $t$-matrix from Eq.~(\ref{eq:3.4}) for 
$p+^{48}$Ca for the
same angular momenta and support points. Those $t$-matrix elements enter the calculation of the
cross section in Fig.~\ref{fig2}.
First we note that for $l=0$ the $p+^{48}$Ca form factors
are quite different from the $n+^{48}$Ca form factors. In addition, they
fall off much slower as function of $p$, a property mainly caused by the short range Coulomb
potential. 

In Fig.~\ref{fig4} we carry out an analogous comparison between the form factors for
the $n+^{208}$Pb and  $p+^{208}$Pb form factors. Here the energies are chosen slightly higher,
since in the $p+^{208}$Pb the form factors at the lowest energies given in Table~\ref{table1}
are very small.
The slow decrease of the $p+^{208}$Pb form factor for the small angular
momentum is  even more pronounced in this case. 

At this point it is crucial to note that in Figs.~\ref{fig3} and~\ref{fig4} we compare two
quite different form factors. For $n+^{48}$Ca and $n+^{208}$Pb
scattering the $t$-matrix elements leading to the form factors are calculated 
as described in Section~\ref{nplusA} using
as basis states in- and out-going plane-wave scattering states.
For $p+^{48}$Ca and $p+^{208}$Pb,
the Coulomb distorted nuclear $t$-matrix elements enter the cross section and lead
to the form factors.
Those Coulomb distorted $t$-matrix elements are evaluated 
in the basis of Coulomb scattering states. Thus, one
should not be surprised that the form factors given in the left and right panels of 
Figs~\ref{fig3} and~\ref{fig4} differ from each other. 

%%%%%%%%%%%%%%%%%%%%%%%%%%%%%%%%%%%%%%%%%%%%%%%%%%%%%%%%%%%%%%%%%%%%%%%%%%%%%%%%%%%%

\section{Coulomb distorted Neutron-Nucleus Form \\ Factors}
\label{cnplusA}

In order to treat charged-particle scattering in momentum space without employing a
screening procedure for the Coulomb force, it is necessary to formulate the scattering
problem in a momentum space Coulomb basis. For proton-nucleus scattering, 
a two-body problem with a repulsive Coulomb force, the Coulomb distorted 
nuclear matrix elements are already derived in this bases, 
as described in the previous Section
and Refs.~\cite{Hlophe:2014soa,Elster:1993dv,Chinn:1991jb}. When moving forward to 
$(d,p)$ reactions, an effective three-body problem with
two charged particles,
one needs to solve generalized Faddeev-AGS equations in Coulomb basis, as was
proposed in Ref.~\cite{Mukhamedzhanov:2012qv}. In order for this approach to be
numerically practical, reliable techniques to calculate expectation values in
this basis must exist. Here we evaluate the neutron-nucleus form factors 
from Section~\ref{nplusA}
in the Coulomb basis to illustrate the feasibility of the approach.

\vspace{9mm}
\begin{figure}[h]
\centerline{\includegraphics[width=10.0cm]{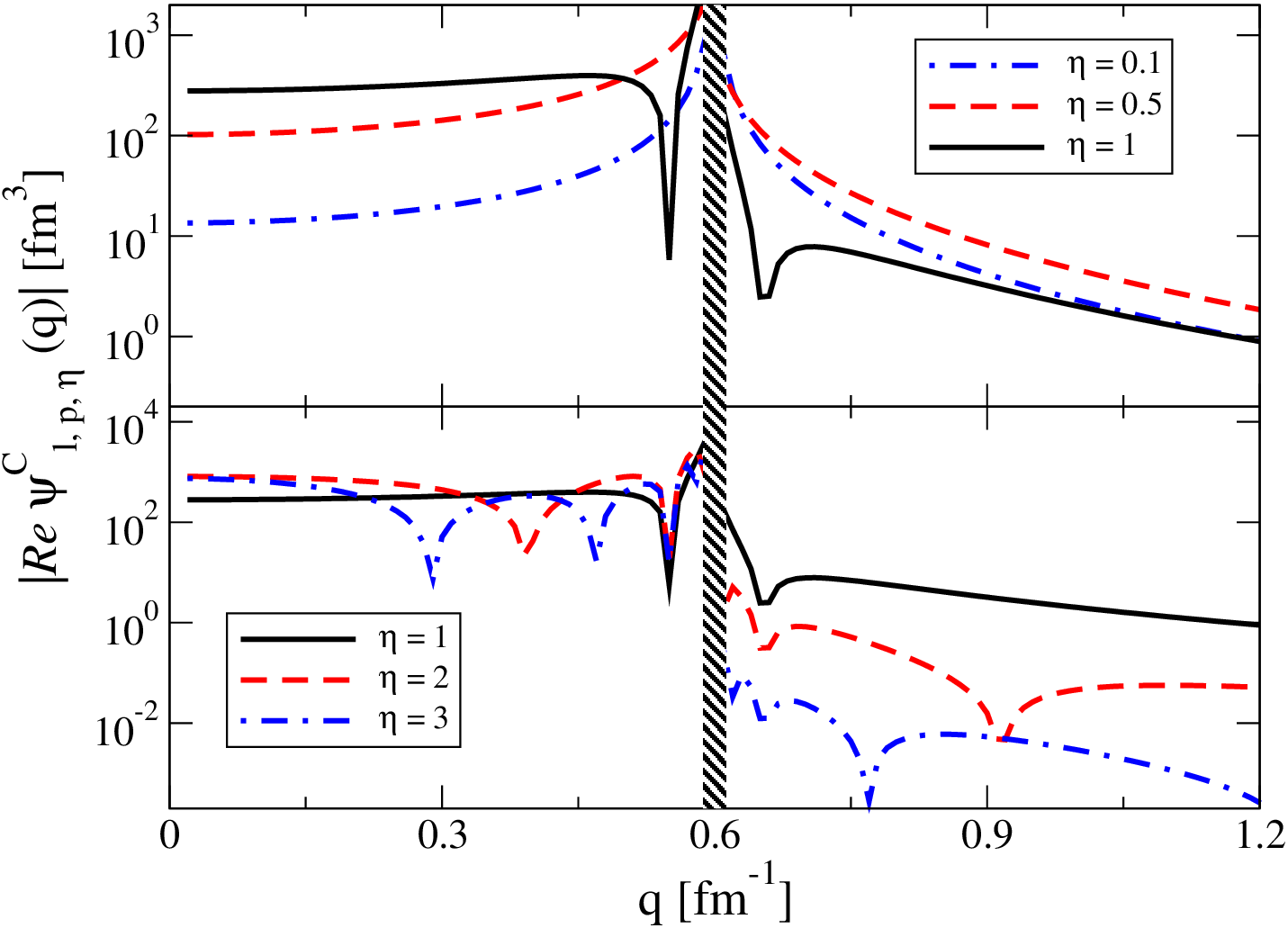}}
\caption{
The absolute value of the real part of
the $l=0$ Coulomb wave function $\psi^C_{l,p,\eta}(p)$ for the external momentum  
$p=0.6$~fm$^{-1}$ and $\eta=0.1, 0.5, 1 $ (upper panel) and $\eta=1, 3, 3$ (lower panel),  
as function of $q$.  The shaded area masks the
function around the singularity at $p \rightarrow q$, where it is highly oscillatory.
}
\label{fig5}
\end{figure}

The starting point is the analytic expression for the Coulomb wave function in momentum
space which, after
a partial wave decomposition, can be written as  (see
\cite{Mukhamedzhanov-private} and
Ref.~\cite{upadhyay:2014}) \\
\begin{equation}
\psi^C_{l,p}(q)\,=\,-\,\frac{2\pi\,e^{\eta\pi/2}}{pq}\,
\lim_{\gamma\to+0}\,\frac{d}{d\gamma} \left\{
\left[\frac{q^2-(p+i\gamma)^2} {2pq}\right]^{i\eta}\,
(\zeta^2-1)^{-i\frac{\eta}{2}}\,Q_l^{i\eta}(\zeta)\right\}\; .
\label{eq:4.1}
\end{equation}
Here, $p$ is the magnitude of the fixed asymptotic momentum and
$\zeta=(p^2+q^2+\gamma^2)/2pq$. The Sommerfeld parameter is given as $\eta=Z_1 Z_2 e^2
\mu/p$
where $Z_1=1$ and $Z_2$ corresponds to the number of protons in the nucleus, and $\mu$ is the
reduced mass of the two-body system under consideration.
The spherical function $Q_l^{i\eta}(\zeta)$ in Eq.~(\ref{eq:4.1}) can be expressed
in terms of hypergeometric functions $_2F_1$ \cite{Dolinskii1966}. However,
care must be taken in its evaluation, since
there are specific limits of validity of the various expansions. Specific difficulties
together with the
expressions implemented in this work are discussed in detail in
Refs.~\cite{upadhyay:2014,coulwavefct}.

In Fig.~\ref{fig5} we display  $l=0$ partial wave Coulomb functions for fixed external
momentum $q=0.6$~fm~$^{-1}$ as function of $p$ for selected values of $\eta$. The
functions exhibit oscillatory singular behavior for $p \rightarrow q$. This region is
indicated in the figure by the shaded band. For values of $\eta \ge 1$ oscillatory
behavior is already present way outside the singular region. It is also worthwhile to
note that once the momentum $p$ is larger than the external momentum $q$, the magnitude
of the Coulomb function falls off by at least an order of magnitude. 

For evaluating the neutron-nucleus form factors in the Coulomb basis,
we start from the separable partial-wave $t$-matrix operator given in Eq.~(\ref{eq:2.8}).
Evaluating its momentum space matrix elements $\langle p|t_l(E)|p'\rangle$ in
a plane-wave basis gives the nuclear form factors
\begin{eqnarray}
\langle p|u| f_{l,k_E}\rangle &=& t_l(p,k_E;E_{k_E}) \equiv u_l(p) \nonumber \\ 
\langle f^*_{l,k_E}|u|p'\rangle &=& t_l(p',k_E;E_{k_E}) \equiv u_l(p'),
\label{eq:4.2}
\end{eqnarray}
where the $t_l(p,k_E;E_{k_E})$ are the half-shell two-body t-matrices obtained as
solution of a
momentum space LS equation with the complex potential $u$. 

The corresponding Coulomb-distorted form factors are obtained by replacing the plane-wave
basis state
by a Coulomb basis state $|\psi^C_{l,p}\rangle$ leading to
\begin{eqnarray}
\langle \psi^C_{l,p} |u|f_{l,k_E}\rangle &=& \int_0^\infty \frac{dq\;
q^2}{2\pi^2} \; u_l(q)  \psi^C_{l,p}(q)^\star  \equiv u^C_l (p) \label{eq:4.3} \\
\langle f^*_{l,k_E}|u| \psi^C_{l,p} \rangle &=& \int_0^\infty \frac{dq\;
q^2}{2\pi^2} \;  u_l(q)\; \psi^C_{l,p}(q) \equiv u^C_l(p)^\dagger  \label{eq:4.3b}
\end{eqnarray}
When $\eta \rightarrow 0$, Eqs.(\ref{eq:4.3}) and  (\ref{eq:4.3b}) tend to
Eq.(\ref{eq:4.2}).
This expression is a generalization of the form introduced in
Ref.~\cite{Mukhamedzhanov:2012qv} to account for complex interactions.

The main challenge in computing the integrals of Eq.~(\ref{eq:4.3}) and (\ref{eq:4.3b})
 is the oscillatory singularity in the integrand for $q=p$,
which is of the form
\begin{equation}
S(q-p) = \lim_{\gamma \to +0} \frac{1}{(q-p+i\gamma)^{1+i\eta}}.
\label{eq:4.4}
\end{equation}
This type of singularity cannot be numerically evaluated by the familiar principal value
subtractions but rather
needs to be treated using the scheme of Gel'fand and Shilov~\cite{gelfand}, as proposed
by \cite{Mukhamedzhanov:2012qv,Dolinskii1966}. The generalization  to the  complex form
factors of our application is given in Ref.~\cite{upadhyay:2014}. The essence of the
Gel'fand and Shilov scheme is to subtract as many terms as needed of the Laurent
expansion in a small region around the pole  so that the oscillations around the pole
become small, and the integral becomes regular. For further details of the calculations
as well as numerical tests we refer to Ref.~\cite{upadhyay:2014}.

\begin{figure}
\centerline{\includegraphics[width=10.0cm]{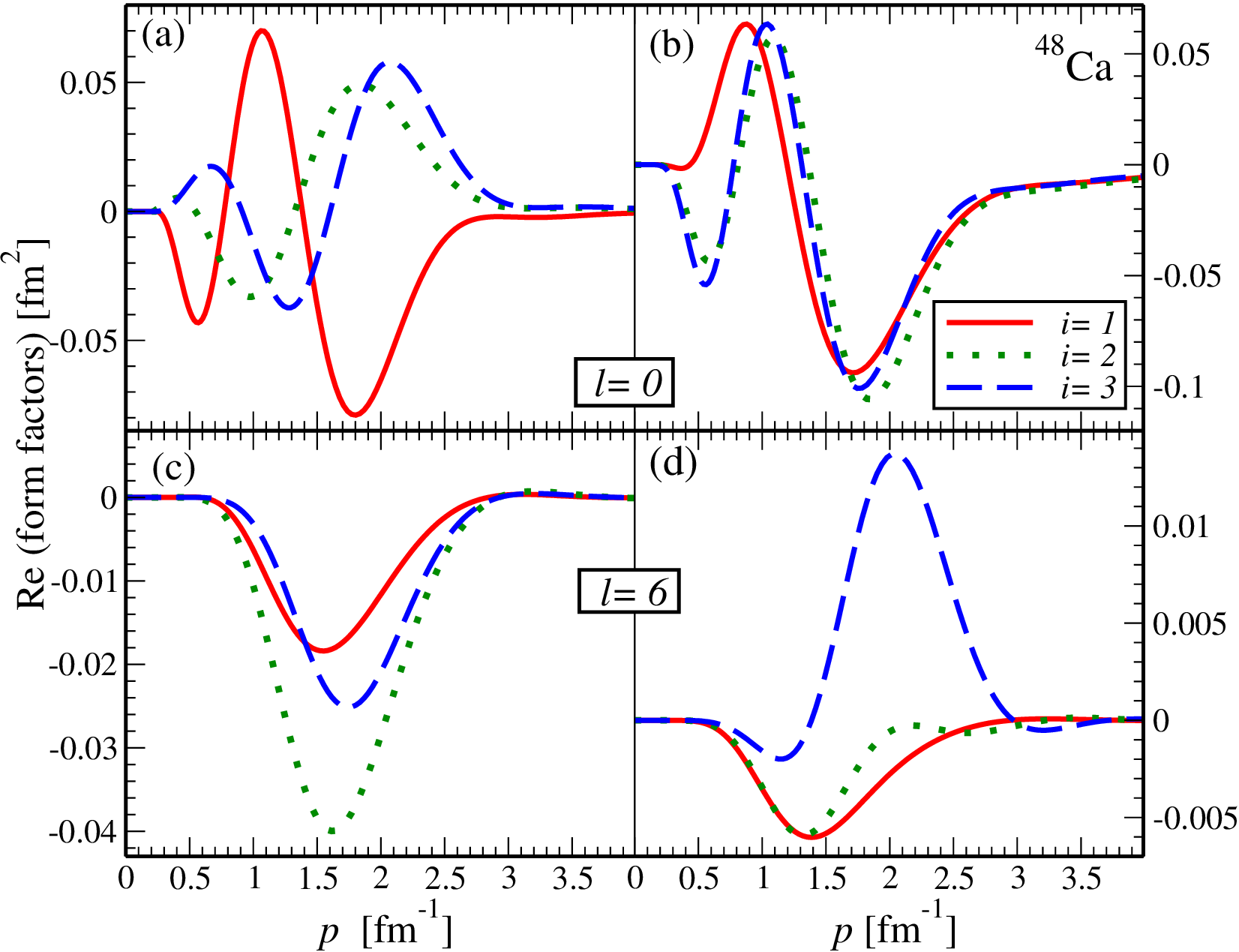}}
\caption{The real parts of the partial wave Coulomb distorted 
neutron form factors for $^{48}$Ca as function of
the momentum $p$ for $l=0$ (a) and $l=6$ (c). 
The form factors are calculated at the energies
indicated in Table~\ref{table1} for the given angular momentum, $1 \equiv$~6~MeV,
$2\equiv$~15~MeV, and $3\equiv$~36~MeV. The real parts of the proton
form factors for $^{48}$Ca as function of the momentum $p$ are given for $l=0$ in (b) and
$l=6$ in (d) for the energies given in Table~\ref{table1}.
}
\label{fig6}
\end{figure}

\begin{figure}
\centerline{\includegraphics[width=10.0cm]{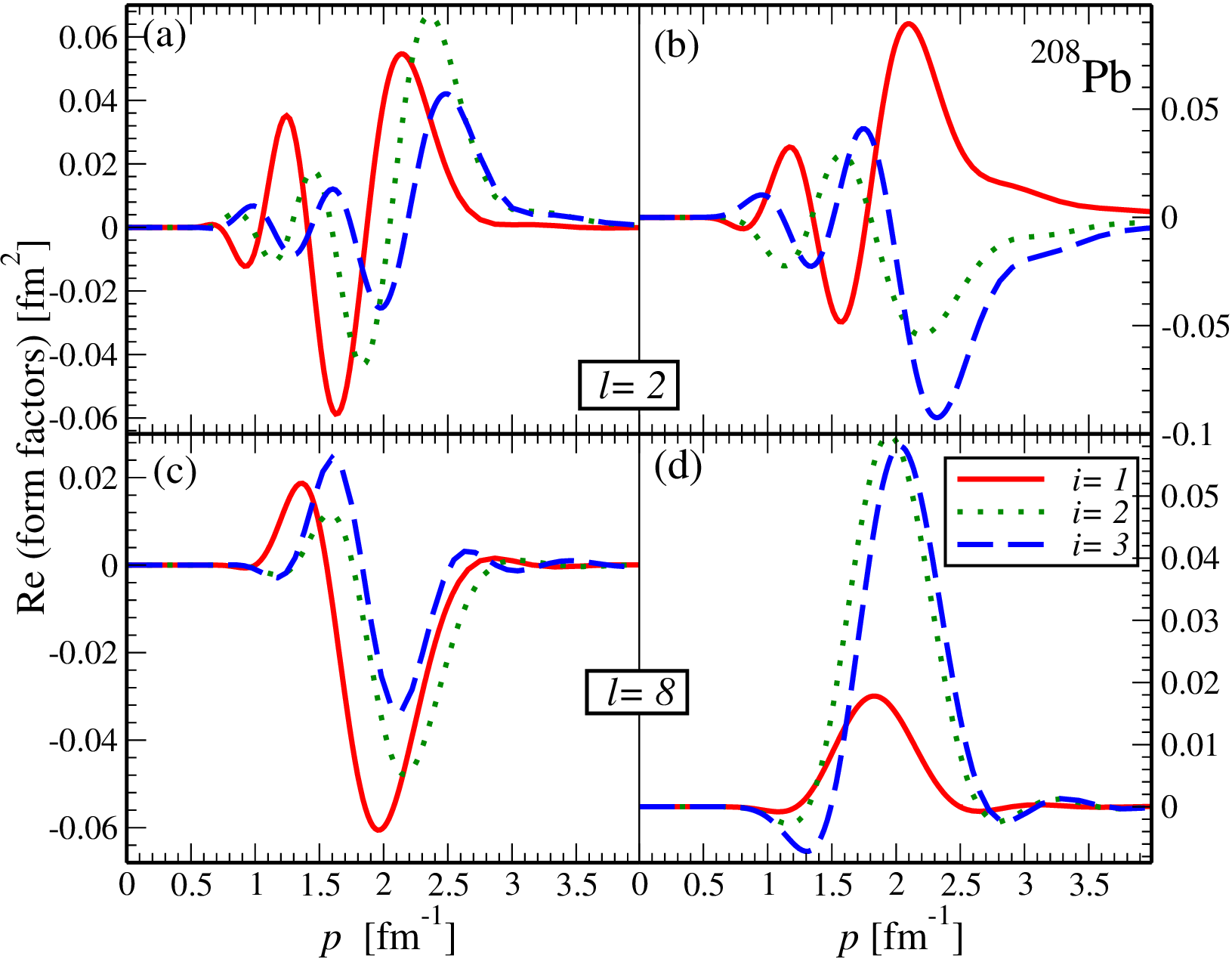}}
\caption{The real parts of the partial wave Coulomb distorted
neutron form factors for $^{208}$Pb as function of
the momentum $p$ for $l=0$ (a) and $l=8$ (c). The form factors are calculated at the
first three energies
indicated in Table~\ref{table1} for the given angular momentum, $1\equiv$~21~MeV,
$2\equiv$~36~MeV, and $3 \equiv$~47~MeV ($l=2$) and 45~MeV ($l=8$). The real parts of
the proton form factors for $^{208}$Pb as function of the momentum $p$ are 
given for $l=2$ in (b) and $l=8$ in (d) for the same energies.
}
\label{fig7}
\end{figure}

In order to illustrate the behavior of Coulomb distorted neutron form factors we show in
Fig.~\ref{fig6} in the left panels the real parts of the Coulomb distorted neutron form
factors of the $n+^{48}$Ca $t$-matrix for $l=0$ (a) and $l=6$ (c) at the same support
points as the plane-wave $n+^{48}$Ca form factors shown in Fig.~\ref{fig3} and the
Coulomb distorted $p+^{48}$Ca form factors shown in the right panels. 
The effect of Coulomb distortions is clearly visible for $l=0$, where the form factor
goes to zero as $p \rightarrow 0$. The figure also shows that the Coulomb distorted
neutron- and proton form factors are quite different. 

In Fig.~\ref{fig7} a similar comparison is shown but for real parts of the Coulomb distorted
$n+^{208}$Pb and $p+^{208}$Pb form factors. Drawing attention to the different scales
for the left and right side panels, we note that the Coulomb distorted $p+^{208}$Pb form
factors do not only differ in shape, but also in magnitude from the Coulomb distorted 
$n+^{208}$Pb form factors. This may not come as a surprise when having in mind that the
Coulomb force is quite strong in heavy nuclei. The comparisons in Figs.~\ref{fig6}
and~\ref{fig7} emphasize the need for a proper introduction of the Coulomb force in the
EST scheme as presented in Section~\ref{pplusA}.  

The realization that the Coulomb distorted neutron-nucleus form factors differ from the
proton-nucleus ones has been already pointed out in Ref.~\cite{Schweiger:1982ic} where 
separable $t$-matrices for proton-proton ($pp$) scattering were considered. 
There the authors used a separable representation in terms of Yukawa functions and
re-adjusted the parameters in the two lowest partial wave to describe the experimentally
extracted $pp$ phase shifts. While such an approach may be viable in the $pp$ system, 
it is not very practical when heavy nuclei are considered, since here many more partial
waves are affected by the Coulomb force.

\begin{figure}
\centerline{\includegraphics[width=10.0cm]{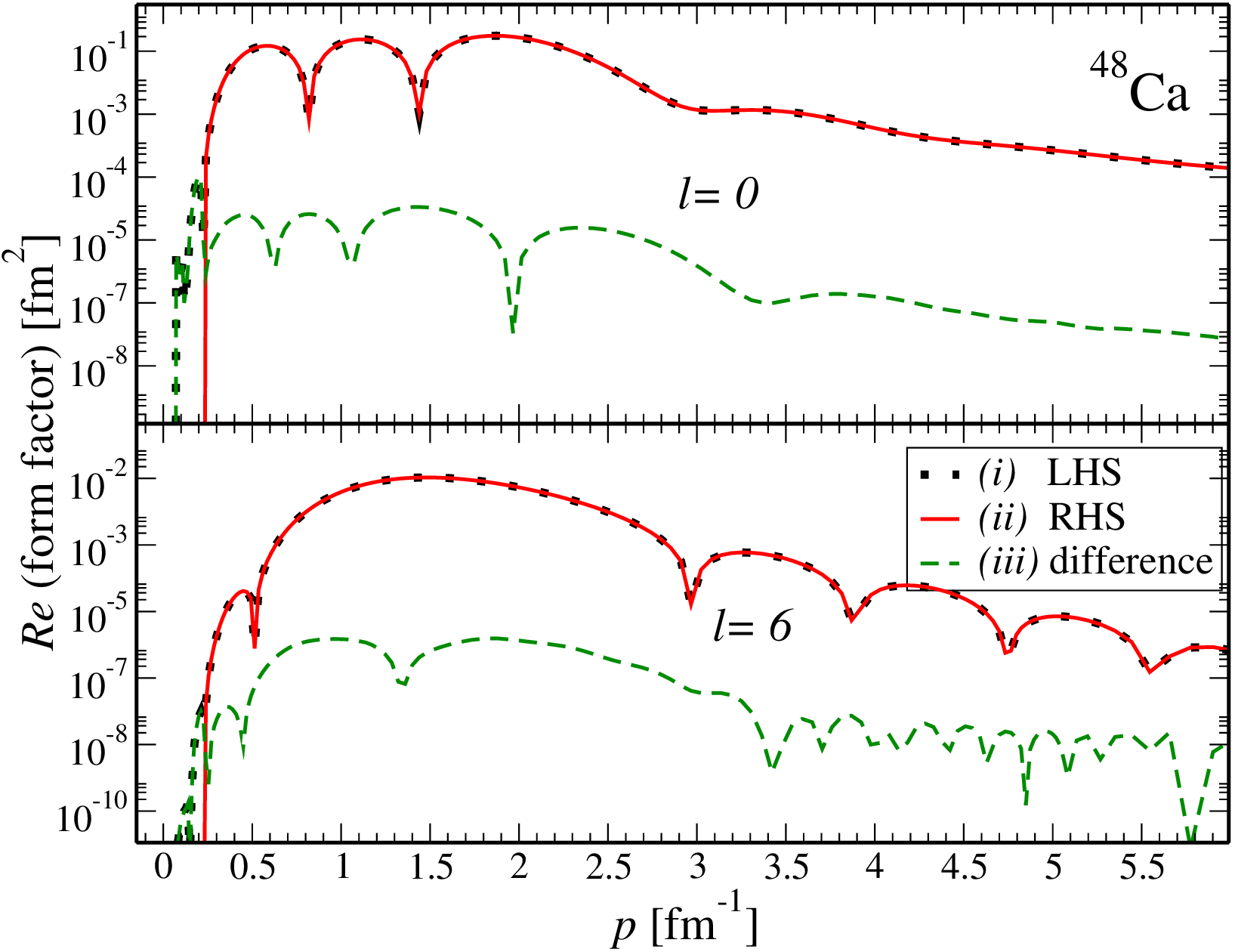}}
\caption{The absolute value of the real part of the partial-wave Coulomb distorted $n+^{48}$Ca for
$l=0$ ($E$=6~MeV) and $l=6$ ($E$=16~MeV) as function of the momentum $p$. 
The dotted lines ($i$) represent the integration over
the Coulomb wave functions, while the solid lines ($ii$) stands for the calculations
 according to right-hand side of Eq.~(\ref{eq:4.6}).
The absolute difference between the two calculation is shown as dashed line ($iii$).
}
\label{fig8}
\end{figure}

Finally, we want to inspect the Coulomb distorted form factor of Eq.~(\ref{eq:4.3}) and consider an
alternative way for its calculation in order to verify the quite involved integration procedure
outlined in this Section and given in detail in Ref.~\cite{upadhyay:2014}. The quantity
$u|f_{l,k_E}\rangle$ satisfies an operator LS equation,
\begin{equation}
u|f_{l,k_E}\rangle = u|k_E\rangle + u g_0(E) u|f_{l,k_E} \rangle,
\label{eq:4.5}
\end{equation}
where $|k_E\rangle$ is the radial part of the solution of the free Hamiltonian at energy $E$ with
angular momentum $l$, and $g_0(E)$ is the free Green's function. Multiplying from the left with the
Coulomb scattering wave function $\psi^c_{l,p}$ gives 
\begin{equation}
\langle \psi^c_{l,p} |u|f_{l,k_E}\rangle = \langle \psi^c_{l,p} |u| k_E\rangle +
\int dp' p'^2 \langle \psi^c_{l,p} |u| p'\rangle \frac{1}{E- E_{p'} + i\epsilon}
  \langle p' |u|f_{l,k_E}\rangle.
\label{eq:4.6}
\end{equation}
The term $\langle p' |u|f_{l,k_E}\rangle = t_l(p',k_E;E_{k_E})$ is the half-shell
$t$-matrix at a support point $E_{k_E}$ already calculated when obtaining the form factors for the
separable representation (see Eq.~(\ref{eq:4.2})). It remains to calculate the driving term, 
which now is given as
\begin{eqnarray}
  \langle  \Phi_{l,p'}^{c} |u| p\rangle 
&=& \frac{2}{\pi}\int\limits_0^\infty \langle\Phi_{l,p'}^{c}|r' \rangle
  \;r'^2 dr' \;\langle r'|u|r\rangle  \;r^2 dr\;\langle r| p\rangle \cr
&=& \frac{2}{\pi p'} \int\limits_0^\infty  dr r^2 dr' r' F_l(\eta',p'r')\;
  \langle r'| u |r \rangle  \;j_l(pr),
\label{eq:4.7}
\end{eqnarray}
which turns for the phenomenological Woods-Saxon potential into 
\begin{equation}
\langle  \Phi_{l,p'}^{c} |u| p\rangle = \frac{2}{\pi p'} \int\limits_0^\infty  dr r 
\; F_l(\eta',p'r)\; u^s(r) \; j_l (pr) .
\label{eq:4.8}
\end{equation}

We now can evaluate the left-hand side (LHS) and the right-hand side (RHS) of Eq.~(\ref{eq:4.6})
independently with two completely different algorithms. This comparison is shown
for two different form factors for $^{48}$Ca. For the $l=0$ the form factor at $E$~=~6~MeV
is shown, for $l=6$ the one at $E$~=~16~MeV. 
The results of both independent calculations indistinguishable in the graph. Thus we show the
absolute difference between the two calculations as dashed line. This shows that our numerical
integration over the momentum-space Coulomb functions together with the Gel'fand-Shilov regularization
is very accurate and can be used without any problem in Faddeev-AGS equations formulated 
in the Coulomb basis when matrix elements in this basis may only be obtained in this fashion.

%%%%%%%%%%%%%%%%%%%%%%%%%%%%%%%%%%%%%%%%%%%%%%%%%%%%%%%%%%%%%%%%%%%%%%%%%%%%%%%%%%%%

\section{Summary and Outlook}
\label{summary}

In a series of steps we developed the input that will serve as a  basis for Faddeev-AGS
three-body calculations of
$(d,p)$ reactions, which will not rely on the screening of the
Coulomb force. To achieve this, Ref.~\cite{Mukhamedzhanov:2012qv} formulated the
Faddeev-AGS equations in the Coulomb basis using separable interactions in the two-body
subsystems. For this ambitious program to have a chance of being successful, the 
interactions in the two-body subsystems, namely the NN and the neutron- and
proton-nucleus systems, need to developed so that they separately describe the observables
of the subsystems. While for the NN interaction separable representations are available,
this is was not the case for the optical potentials describing the nucleon-nucleus
interactions. Furthermore, those interactions in the subsystems need to be available in
the Coulomb basis.

We developed separable representations of phenomenological optical
potentials of Woods-Saxon type for neutrons and protons. First we concentrated on
neutron-nucleus optical potentials and generalized the Ernst-Shakin-Thaler (EST)
scheme~\cite{Ernst:1973zzb} so that it can be applied to complex
potentials~\cite{Hlophe:2013xca}. In order to consider proton-nucleus optical potentials,
we further extended the EST scheme so that it can be applied to the scattering of charged
particles with a repulsive Coulomb force~\cite{Hlophe:2014soa}.  
While the extension of the EST scheme to charged particles led to a separable 
proton-nucleus $t$-matrix in the Coulomb basis, we had to develop  methods to
reliably compute Coulomb distorted neutron-nucleus $t$-matrix
elements~\cite{upadhyay:2014}. Here we also show explicitly that those  calculations
can be carried out numerically very accurately by calculating them within two independent
schemes.

Our results demonstrate, that our  separable representations
 reproduce standard coordinate space
calculations of neutron and proton scattering cross sections very well, and that we are able to 
accurately compute the integrals leading to the Coulomb distorted form factors. Now that
these challenging form factors have been obtained, they can be introduced into the
Faddeev-AGS equations to solve the three-body problem without resorting to screening. Our
expectation is that solutions to the Faddeev-AGS equations written in the
Coulomb-distorted basis can be obtained for a large variety of  $n+p+A$ systems, without
a limitation on the charge of the target.
From those solutions,  observables for $(d,p)$
transfer reactions should be readily calculated. Work along these lines is in
progress.

%%%%%%%%%%%%%%%%%%%%%%%%%%%%%%%%%%%%%%%%%%%%%%%%%%%%%%%%%%%%%%%%%%%%%%%%%%%%%%%%%%%%

\vspace{5mm}
\begin{center}{\bf Acknowledgments} \end{center}
\vspace{-1mm}
This material is based on work  in part supported
by the U.~S.  Department of Energy, Office of Science of Nuclear Physics
under  program No. DE-SC0004084 and  DE-SC0004087 (TORUS Collaboration), under contracts
DE-FG52-08NA28552  with
Michigan State University, DE-FG02-93ER40756 with Ohio University;  by  Lawrence Livermore
National Laboratory under Contract DE-AC52-07NA27344 and the U.T. Battelle LLC Contract
DE-AC0500OR22725.
F.M. Nunes  acknowledges support from the National Science Foundation
under grant PHY-0800026. This research used resources of
the National Energy Research Scientific Computing Center, which is supported by the Office of
Science of the U.S. Department of Energy under Contract No. DE-AC02-05CH11231.

%%%%%%%%%%%%%%%%%%%%%%%%%%%%%%%%%%%%%%%%%%%%%%%%%%%%%%%%%%%%%%%%%%%%

%\bibliographystyle{unsrt}
%\bibliographystyle{h-physrev4}
\bibliographystyle{h-physrev5}

\bibliography{coulomb}

\end{document}